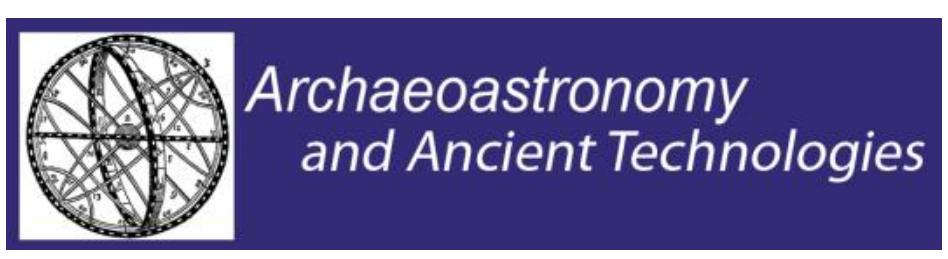

# Reconstruction of Ancient Egyptian Sundials

**Larisa N. Vodolazhskaya**

Department of Space Physics, Southern Federal University (SFU), Rostov-on-Don, Russian Federation; E-mails: larisavodol@aaatec.org, larisavodol@gmail.com

**Abstract**

The article presents the results of a study of the design features of vertical and L-shaped ancient Egyptian sundials, developed their models, on the basis of which the reconstruction of the sundial was carried out. An original scheme of a simple method for a fairly accurate time measurement with their help is also proposed. The model proposed by us, which describes a vertical sundial, is a vertical sundial with an inclined gnomon that takes into account the latitude of the area. The model proposed by us is characterized by marking the hour lines from 6 to 12 o'clock every hour. From 12:00 a shift is made in the marking of the hour lines for half an hour, and then the marking is repeated every hour. As a consequence of the reconstruction of the vertical sundial, we have developed and proposed a model that describes the design features and operation of two types of L-shaped sundial. They had to work in conjunction with an inclined gnomon, like a vertical sundial, or directly with a vertical sundial. In this case, the L-shaped sundial could complement the vertical sundial by providing the ability to read hour markers and interpret the readings of the vertical sundial, as there were no inscriptions on the vertical sundial. The article also provides a decoding of the inscription from the tomb of Seti I, which has long intrigued researchers. It is proved that the inscription contains the lengths of the intervals between adjacent marks of the L-shaped sundial of the second type, where the first interval corresponds not to one hour, but to half an hour.

## Introduction

Technologies of ancient Egypt, including measurement of time, reached a high level. In ancient Egypt, there were clepsydras and sundials, which had a different design: L-shaped, stepped, vertical sundials. Examples of such tools are stored in the Cairo and Berlin museums.

The earliest written evidence of the existence of a sundial in Ancient Egypt dates from the reign of Pharaoh Thutmose III (1521-1473 BC). This pharaoh made several dozen campaigns both in Asia and in Nubia. In the description of one battle in the Manedo Gorge, which he fought during his first campaign in Asia, there is a mention that the army set out at noon, when the shadow of the sun "turns" [1], [2], [3]. It is believed that this moment could only be determined by the sundial.

However, if the very first mention of the ancient Egyptian sundial is still associated with the name of Thutmose III, then, as the Egyptologist D. G. Breasted rightly points out [4], this does not mean that there were no sundials in ancient Egypt before him.

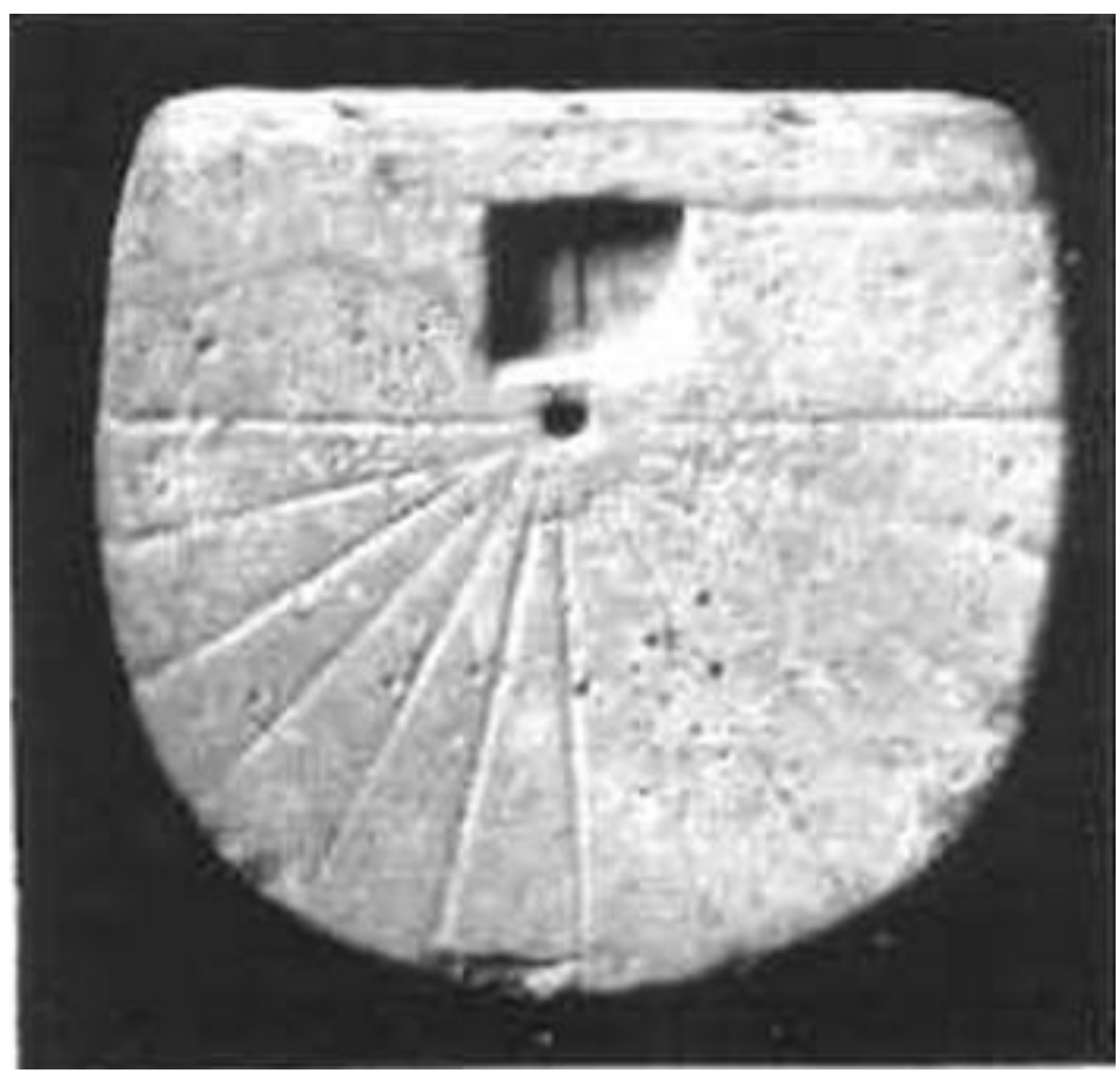

**Figure 1.** Egyptian sundial (XIII century BC).

Also in the tomb of Seti I (1300 BC) there is an image of a sundial [7]. Currently, there are several examples of Egyptian sundial stored in various museums around the world. The Egyptian Museum in Berlin has a green slate sundial dating from the reign of Thutmose III and a Faiyum sundial dating from 1000–600 BC [8]. Another sundial is known from the reign of Pharaoh Merneft (1258-1239 BC), which were found in Palestine (Fig. 1) [9]. On a sundial from Palestine, hour lines emanating from the center are plotted on a flat surface. In ancient times, this sundial was used to determine time, as well as modern sundial, with the help of the direction of the shadow.

It is believed that the division into hours first appeared in Egypt. Already from 2100 BC. The Egyptian priests divided the day into 24 hours [5]. However, the length of the hour constantly changed throughout the year and was equal to 1/12 of the time from sunrise to sunset or from sunset to sunrise. Thus, the duration of an hour fluctuated depending on latitude and time of year [6]. According to the vertical sundial in ancient Egypt, they could more or less accurately determine the hours of the day only during the spring and autumn equinoxes; at other times they showed the hours of the day approximately.

In 2013, during archaeological excavations in the Valley of the Kings, an expedition from the University of Basel (Switzerland) led by S. Bickel and E. Paulin-Grothe discovered a limestone tile with a hole and fan-shaped lines diverging from it ( Fig. 2). The authors of the find suggested that it was a sundial.

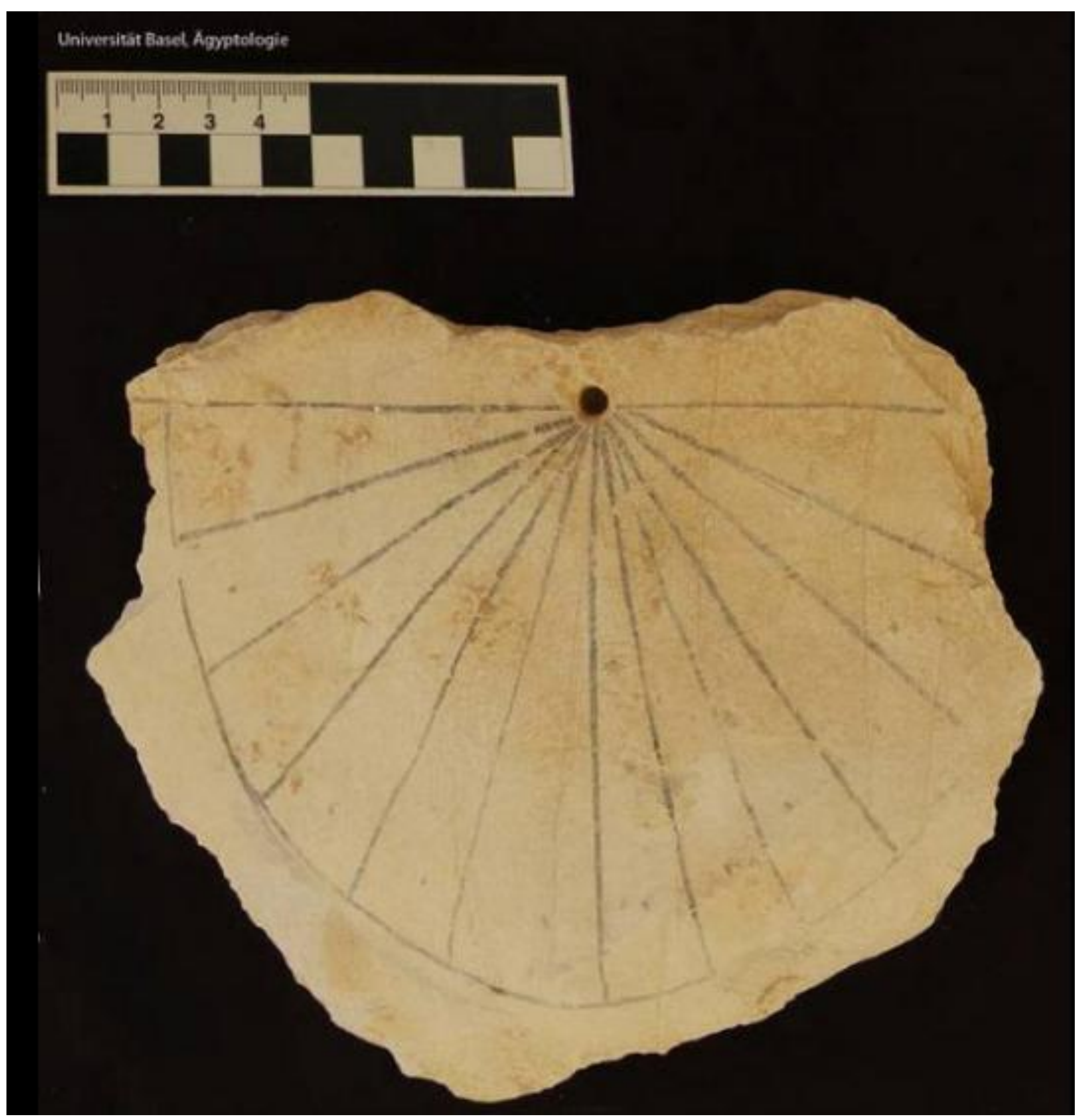

**Figure 2.** Limestone tiles discovered during archaeological excavations in the Valley of the Kings[1]

The sundial was a semicircular piece of local limestone. The front is relatively flat, while the back is very uneven. Dimensions are 15,5 cm high, 17,5 cm wide and a maximum depth of 3,6 cm. The upper right edge of the image is broken off. The dial is painted black on the front. It consists of a horizontal line, in the middle of which a hole about 10 mm deep and about 6 mm in diameter is drilled. A gnomon was attached here, which was not found. It could be a wooden, bronze or lead rod.

In the case of a vertical sundial, the gnomon could be installed either horizontally or obliquely along the axis of rotation of the Earth. The horizontal line on the tile marks the hour lines for 6 am (left) and 6 pm (right). The midday line (12 o'clock) runs almost at right angles (91° or 89°) to the horizontal line. Five more hour lines are marked between them, from the hole to the edge of the tile. A dot is drawn in each of the sectors. Behind the left half of the bounding line, you can see the first erased line. The beginning of the incomplete line is also visible on the first oblique on the right (17:00) [10].

---

[1] http://aegyptologie.unibas.ch/forschung/projekte/university-of-basel-kings-valley-project/report-2013/ "Preliminary Report on the Work carried out during the season 2013", fig. 5 (accessed on 01.08.2014)

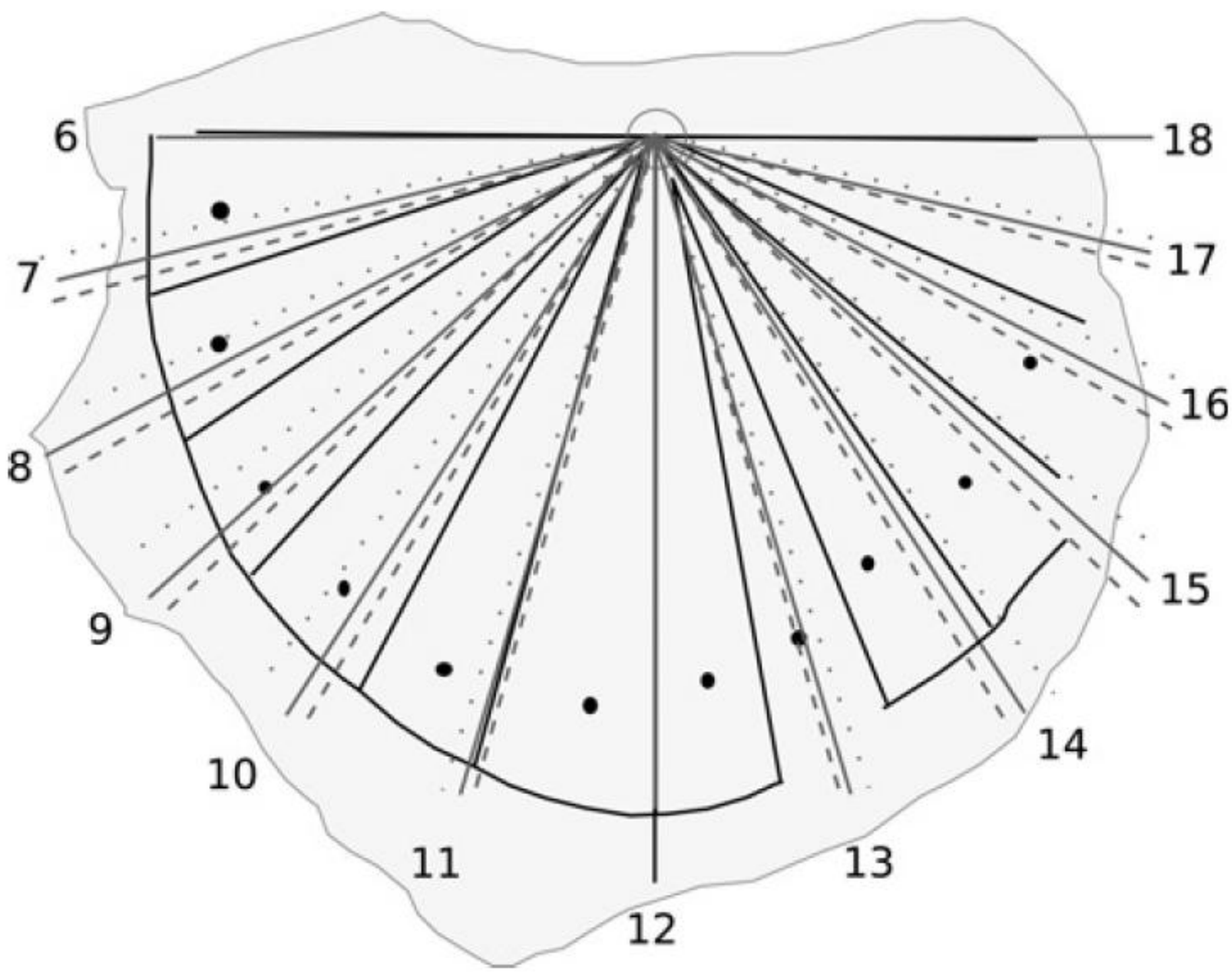

**Figure 3.** Limestone tile from the Valley of the Kings, drawing with applied hour lines calculated for a vertical sundial with a horizontal gnomon. The noon line was chosen as the zero point. The lines on the limestone tile are shown in the drawing as black solid lines, the hour lines calculated for the equinox are shown as gray solid lines, for the summer solstice - gray dotted lines, for the winter solstice - gray lines of dots [11].

In 2014, a study by S. Bickel and R. Gautschy was published, in which they note that the greatest coincidence with the lines on the tile is given by the calculated hour lines of the vertical sundial, oriented to the south, and an inclined gnomon oriented along the Earth's axis of rotation. At the same time, the researchers noted that a good match between the calculated lines and those drawn on the tile is observed only in the range from 7 to 12 noon.

The authors of the study were unable to explain the poor match between 12 and 17 o'clock and concluded that the lines on the tile are approximate and possibly belong to another type of clock - a vertical sundial with a horizontal gnomon. However, the hour lines calculated by them for this type of watch also poorly coincided with the lines on the tile (Fig. 3) [12].

The aim of the study presented in this article was to reconstruct and develop a sundial model that describes the found sundial from the Valley of the Kings with greater accuracy than the previously proposed model, as well as the reconstruction of methods for measuring time using vertical and L-shaped sundials in ancient Egypt.

**Vertical sundial from the Valley of the Kings**

In the process of research, we analyzed the markings of the sundial from the Valley of the Kings. Having carried out calculations for horizontal and vertical sundials, we came to the conclusion that the markings of the found clocks are most similar to those of vertical sundial with an inclined gnomon, especially in the range from 6 to 12 hours. When studying the lines depicted on the limestone tile, one can notice the asymmetry of the sectors of the semicircle relative to the central line (the assumed noon line). If these lines on the tile are hourly, then the central line will be the 12 o'clock line, and the sectors symmetrical with respect to it, formed by the drawn lines, must have equal angular dimensions. As part of this study, in order to explain the asymmetry of the sectors, a hypothesis was put forward that hour lines could be plotted not only for whole hours, but also for time intervals greater or less than an hour by half an hour. The fact that half an hour could be measured and taken into account is indicated by dots drawn approximately in the middle of each sector between the lines on the tile (Fig. 2, 3). To test the hypothesis, hourly angles were calculated every half hour using the standard formula 1 for a vertical sundial with an inclined gnomon [13]:

$$H^{/} = arctg(cos\,\varphi \cdot tgH), \tag{1}$$

Where $H = 15 \cdot (t - 12)$ - the hour angle of the Sun (for noon $t$=12, $H$=0º), $t$ - the time (hours), $H^{/}$ - the angle formed by the hour line with the noon line, $\varphi$ - the latitude (Fig. 4).

The results of calculating the values of hour angles in the range from 6 to 18 hours for the geographical latitude of the Valley of the Kings (Egypt) φ = 25°44' *N* are presented in tables 1 and 2. Swiss researchers measured and calculated hour angles relative to the six o'clock hour line. In the table, the hour angles are recalculated relative to the noon line.

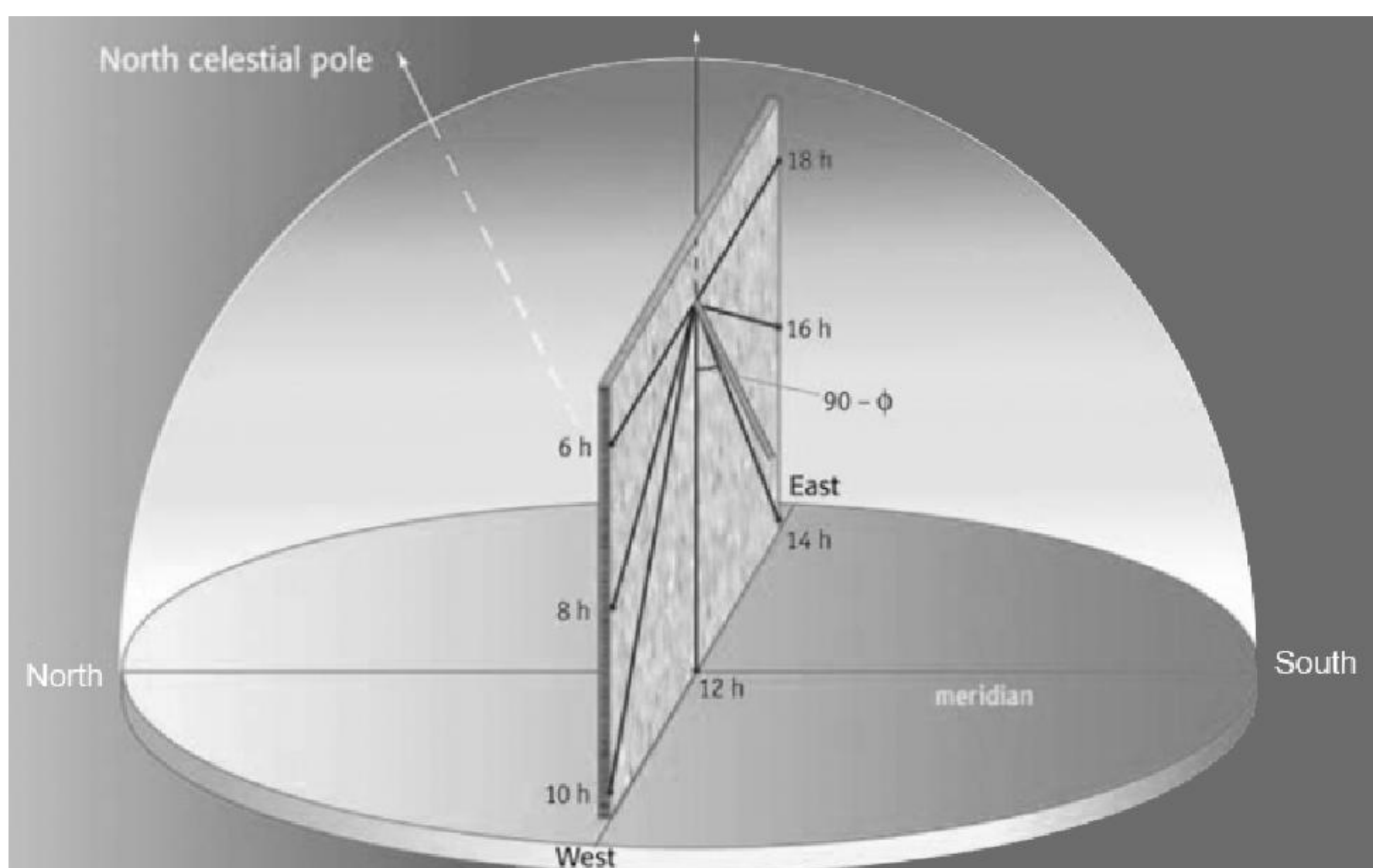

**Figure 4.** Scheme illustrating the principle of operation of a vertical sundial with an inclined gnomon [14].

The average deviation of the calculated hourly angles from the measured ones in the range from 7 to 12 hours is ≈0,8º (Table 1). In the range from 12:00 to 17:00, the average deviation of the calculated hour angles from the measured ones is ≈5,7º (for whole hours), and ≈1º for half an hour. The value of the average deviation of the calculated hour angles from the measured ones, taking into account the shift by half an hour after noon, is close to the value of the average deviation before noon without a shift.

On the image of a tile from the Valley of the Kings, in accordance with the calculations, hour lines were applied corresponding to whole hours in the range from 6 to 12 hours and hour lines corresponding to half hours in the range from 12,5 to 16,5 hours (Fig. 5). When drawing lines on a photograph, the vertices of the calculated hour lines were aligned with the vertices of the corresponding lines on the photograph of the tile.

**Table 1.** Hour angles of a vertical sundial with an inclined gnomon before noon for latitude $25^0 44^/$ N (relative to the noon line): $H$ - the hour angle of the Sun, $H^/$ - the hour angle of the sundial calculated by formula 1, $H_f$ - the measured hour angle on the tile from the Valley of the Kings [15], $H^{//}$ - the hour angle of the sundial, calculated for the equinoxes (recalculated relative to the noon line) [16], $t$ - time.

| | *t, (hour)* | | | | | | | | | | | | |
|---|---|---|---|---|---|---|---|---|---|---|---|---|---|
| | 6,0 | 6,5 | 7,0 | 7,5 | 8,0 | 8,5 | 9,0 | 9,5 | 10,0 | 10,5 | 11,0 | 11,5 | 12,0 |
| *H, (º)* | -90,0 | -82,5 | -75,0 | -67,5 | -60,0 | -52,5 | -45,0 | -37,5 | -30,0 | -22,5 | -15,0 | -7,5 | 0,0 |
| *$H_f$, (º)* | - | - | -73,0 | - | -57,0 | - | -43,0 | - | -30,0 | - | -14,0 | - | 0,0 |
| *$H^/$, (º)* | -90,0 | -81,7 | -73,4 | -65,3 | -57,3 | -49,6 | -42,0 | -34,7 | -27,5 | -20,5 | -13,6 | -6,8 | 0,0 |
| *$H^{//}$, (º)* | - | - | -73,5 | - | -57,5 | - | -42,0 | - | -27,5 | - | -13,5 | - | 0,0 |
| $\|H_f - H^/\|$ | - | - | 0,4 | - | 0,3 | - | 1,0 | - | 2,5 | - | 0,4 | - | 0,0 |
| $\|H_f - H^{//}\|$ | - | - | 0,5 | - | 0,5 | - | 1,0 | - | 2,5 | - | 0,5 | - | 0,0 |

**Table 2.** Hour angles of a vertical sundial with an inclined gnomon after noon for latitude $25^0$ $44^/$ N (relative to the noon line): $H$ - the hour angle of the Sun, $H^/$ - the hour angle of the sundial calculated by formula 1, $H_f$ - the measured hour angle on the tile from the Valley of the Kings, $H^{//}$ - the hour angle of the sundial, calculated for the equinoxes (recalculated relative to the noon line), $t$ - time.

| | *t, (hour)* | | | | | | | | | | | | |
|---|---|---|---|---|---|---|---|---|---|---|---|---|---|
| | 12,0 | 12,5 | 13,0 | 13,5 | 14,0 | 14,5 | 15,0 | 15,5 | 16,0 | 16,5 | 17,0 | 17,5 | 18,0 |
| *H, (º)* | 0,0 | 7,5 | 15,0 | 22,5 | 30,0 | 37,5 | 45,0 | 52,5 | 60,0 | 67,5 | 75,0 | 82,5 | 90,0 |
| *$H_f$, (º)* | 0,0 | 10,0 | 10,0 | 21,0 | 21,0 | 34,0 | 34,0 | 49,0 | 49,0 | 66,0 | 66,0 | - | - |
| *$H^/$, (º)* | 0,0 | 6,8 | 13,6 | 20,5 | 27,5 | 34,7 | 42,0 | 49,6 | 57,3 | 65,3 | 73,4 | 81,7 | 90,0 |
| *$H^{//}$, (º)* | 0,0 | - | 13,5 | - | 27,5 | - | 42,0 | - | 57,5 | - | 73,5 | - | - |
| $\|H_f - H^/\|$ | 0,0 | 3,2 | - | 0,5 | - | 0,7 | - | 0,6 | - | 0,7 | - | - | - |
| $\|H_f - H^{//}\|$ | 0,0 | - | 3,5 | - | 6,5 | - | 8 | - | 8,5 | - | 7,5 | - | - |

It should also be noted that the hour line at 12,5 o'clock is rather carelessly applied to the tile. It is possible that the time from 12 to 13,5 hours corresponded to the midday rest of the workers, next to whose stone dwelling this sundial was discovered. It seems that after the afternoon rest, the working time, just as before noon, was divided into whole hours. At the same time, the first working hour began at 13,5 hours, and each subsequent one - exactly one hour later. At 16,5 o'clock the working day most likely ended, since there are no hour lines or dot marks on the tile after 16,5 to 18 o'clock.

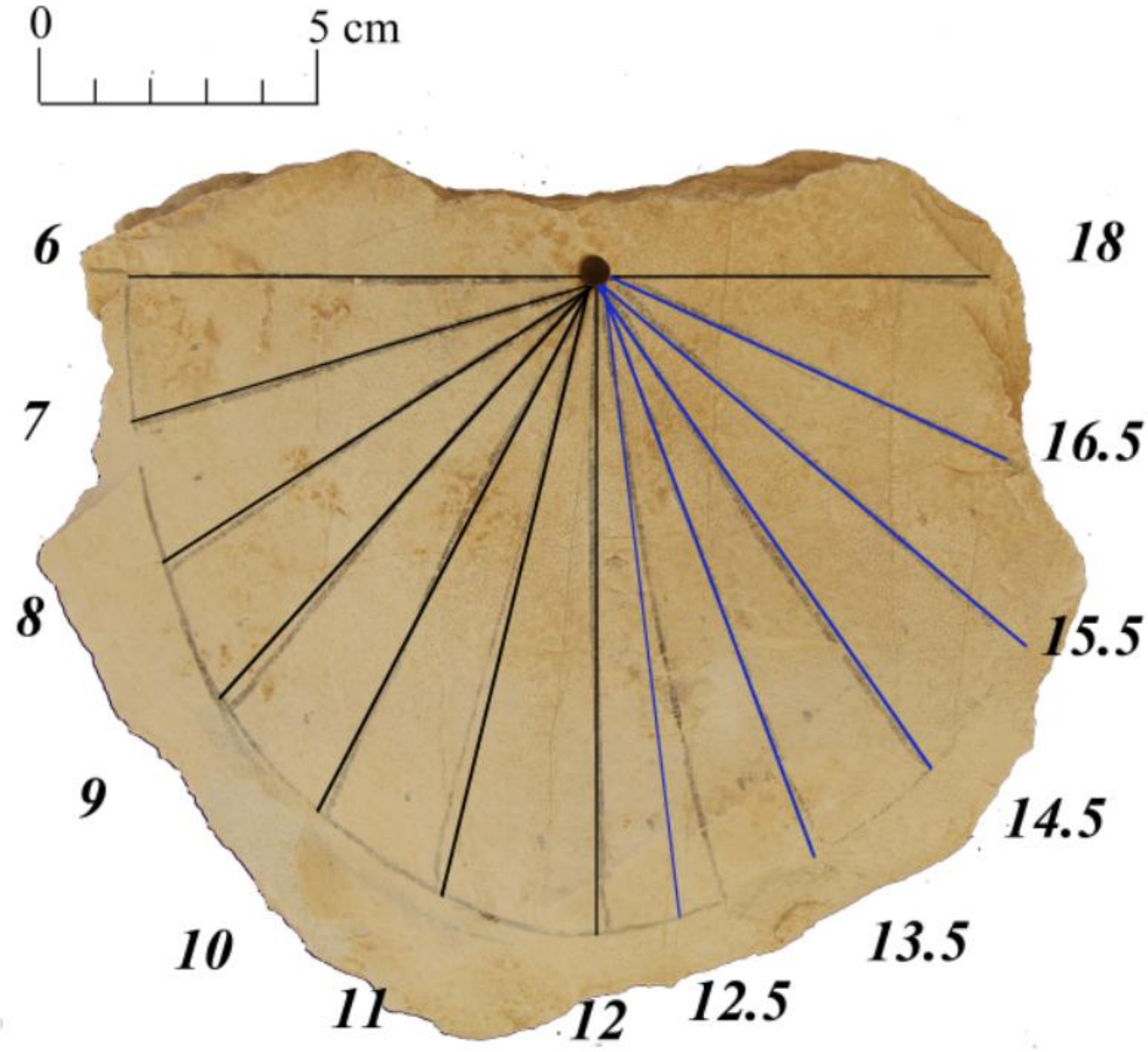

**Figure 5.** Limestone slab from the Valley of the Kings, photograph with hour lines applied to it, calculated hour lines before noon (black lines) and calculated hour lines with a shift of half an hour after noon (blue lines).

The time of noon has been singled out and associated with rest and eating in many traditions. So, for example, it is known that in Ancient Greece the working day began at dawn and lasted until noon, which marked the end of working hours (Anth. Pa1., X, 43 - Anthologia Palatina). Lunch time in Ancient Rome (Mart., IV, 8 - Marcus Valerius Martialis), fell on a period close to noon [17]. In many countries with a hot climate, a midday rest - a siesta - is still widespread. Perhaps the markings of the sundial found in the Valley of the Kings is one of the oldest evidence of the existence of this tradition as far back as the era of Ancient Egypt.

**L-shaped sundial**

The image of a sundial in the form of an L-shaped bar from the tomb of Seti I is well known (Fig. 6).

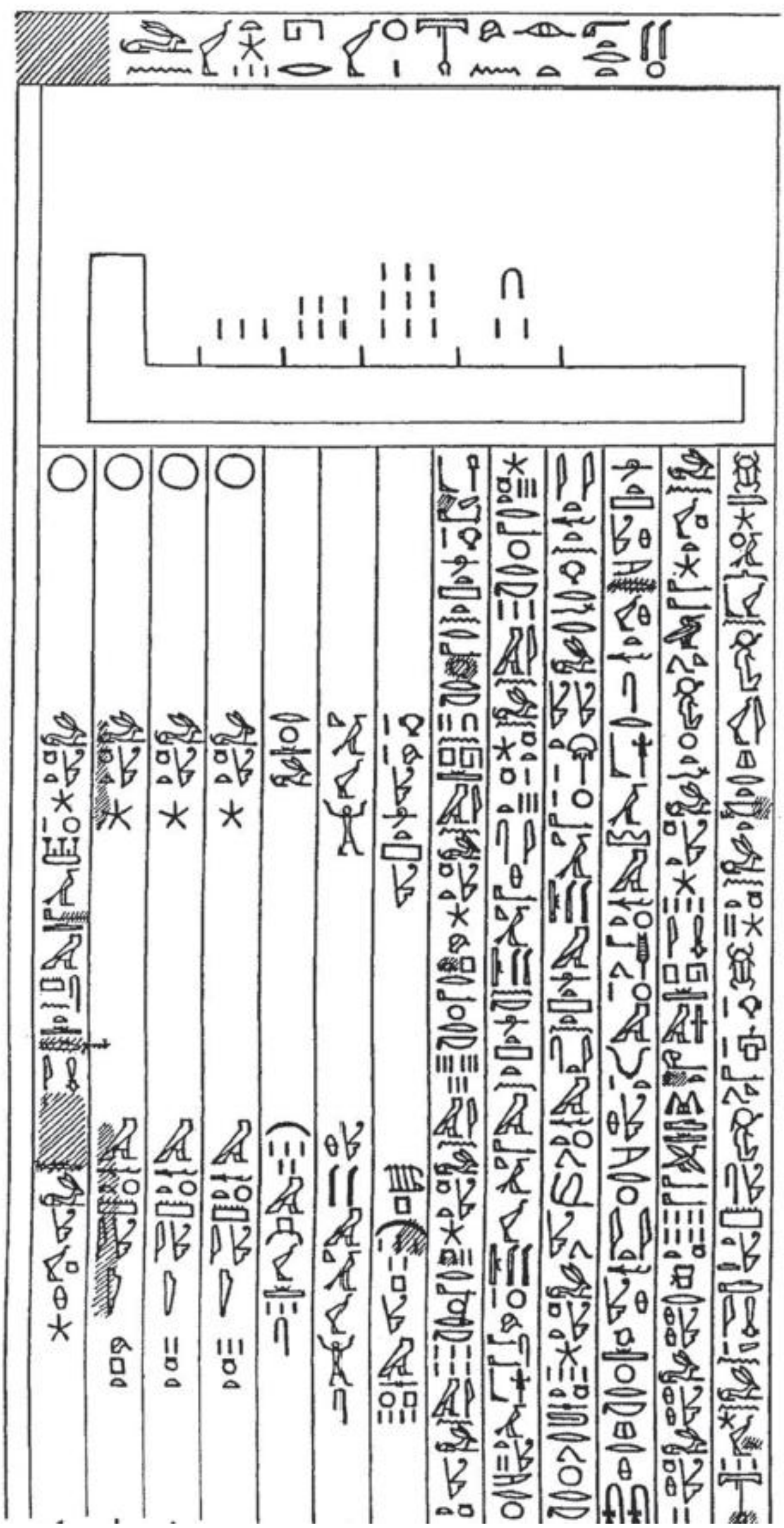

**Figure 6.** Image of a L-shaped sundial from the tomb of Seti I at Abydos [18]

The L-shaped sundial made of green slate dating from the reign of Thutmose III is kept in the Berlin Egyptian Museum under accession number 19744, and a Faiyum sundial dating from 1000-600 BC, are stored under inventory number 19743 (Fig. 7) [19].

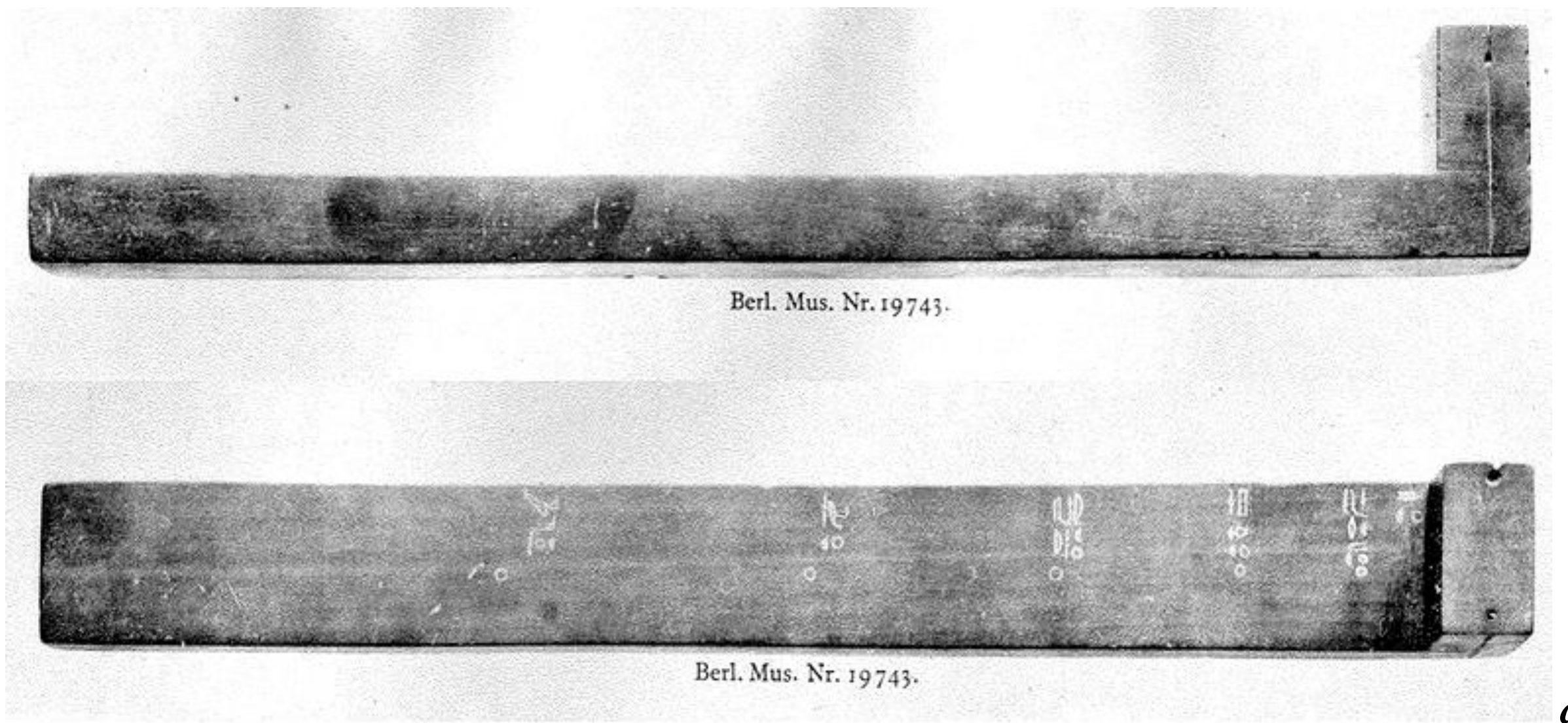

***a***

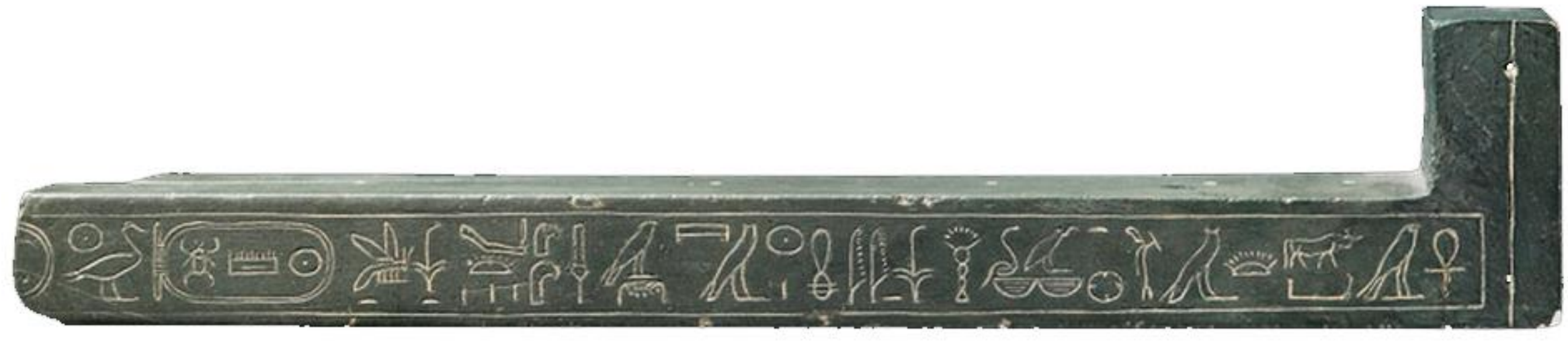

***b***

**Figure 7.** Sundial with a linear scale: a - L-shaped sundial with inventory number 19743[2], b - L-shaped sundial with inventory number 19744[3].

Also quite well known in the scientific world is a fragment of a papyrus papyrus from Tanis of the Roman time, on which in the upper part of the drawing there are lines resembling the hour lines of a vertical clock, and under them, presumably, a fragment of an L-shaped sundial is depicted (Fig. 8a) [20]. The papyrus fragment with the upper part of the L-shaped sundial and the center of the fan-shaped divergent lines is unfortunately missing.

As part of this study, a reconstruction of a more complete image was proposed, which is illustrated in Figure 8b. At the bottom of the figure, in gray, an L-shaped sundial is depicted, as in Figure 7, without any additional crossbar or bar. In the upper part of the figure, at the point

[2] http://members.aon.at/sundials/berlin-egypte.htm (accessed on 01.08.2014)

[3] http://www.aegyptisches-museum-berlin-verein.de/c31.php (accessed on 01.08.2014)

where the fan-shaped lines converge, the place of attachment of the gnomon is assumed, and the fan-shaped lines are the hour lines of the vertical sundial, similar to the lines on the tiles from the Valley of the Kings.

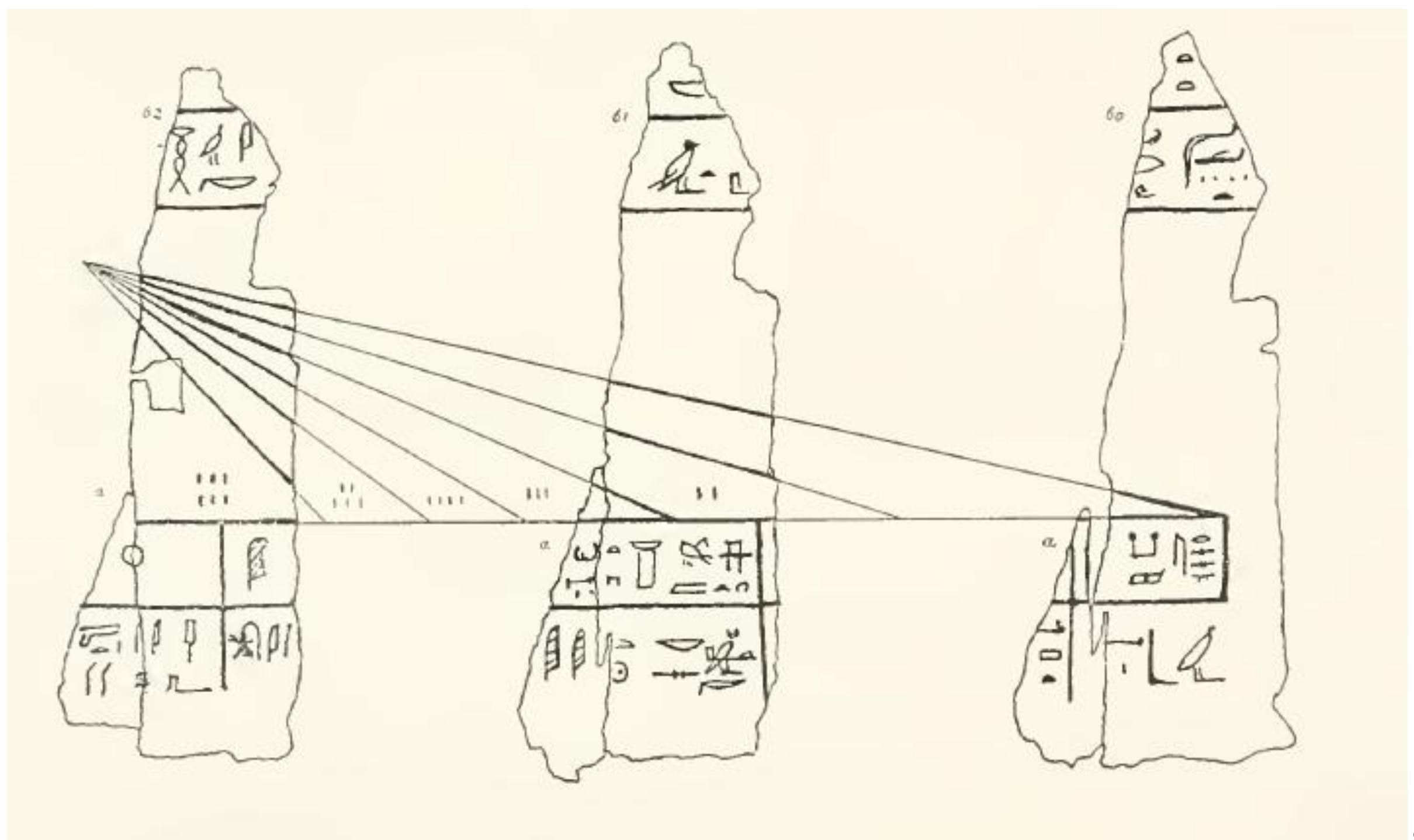

***a***

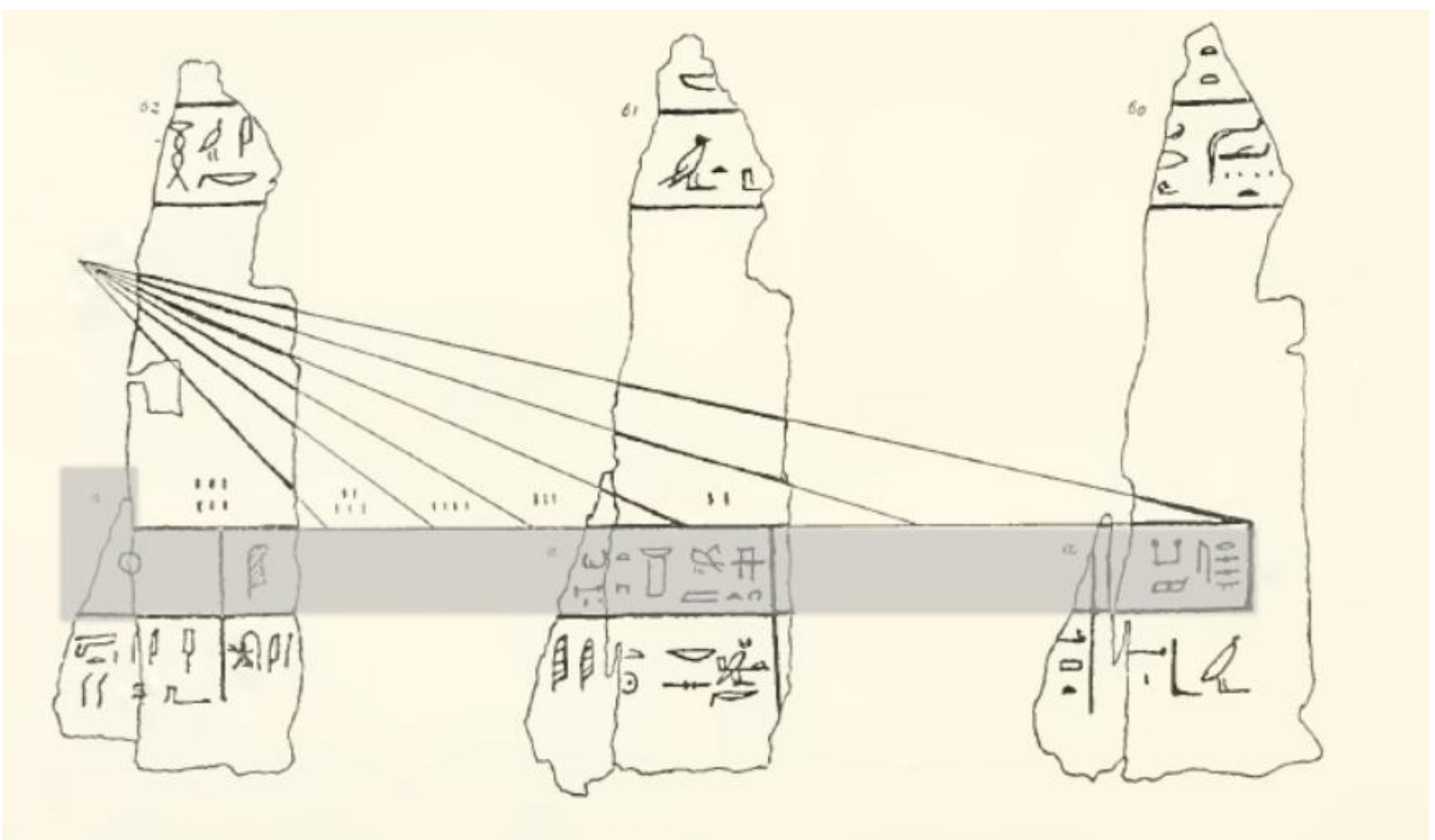

***b***

**Figure 8.** Papyrus fragment, which supposedly depicts a vertical sundial (partial reconstruction by Flinders Petrie [21]): a - drawing of a papyrus fragment, b - reconstruction of the image of an L-shaped sundial (highlighted in gray) on the same papyrus fragment.

Based on this interpretation of the images, it can be assumed that a vertical sundial was used in conjunction with an L-shaped sundial. To test this assumption, we calculated the relative lengths of segments $a_i$ (ratios $a_i/a_1$) formed by the intersection of adjacent hour lines of a vertical sundial with an inclined gnomon in the range from 8 to 11 o'clock with a line perpendicular to the noon line, according to formula 2 (Fig. 9 ):

$$a_i/a_1 = \left(tg(H_i^/) - tg(H_{i-1}^/)\right) / tg(H_1^/) \qquad (2)$$

where $H_i^/$ - hour angle between the noon line ($H_0^/=0$) and the *i*-th hour line, calculated by formula 1; *i* - the number of the hour line, where *i*=1 for the hour line at 11 o'clock, and *i*=4 for 8 o'clock; *ai* - the length of the segment bounded by (*i*-1)-th and *i*-th hour lines, corresponding to the *i*-th hour line.

The results of calculations according to formula 2 are presented in table 3.

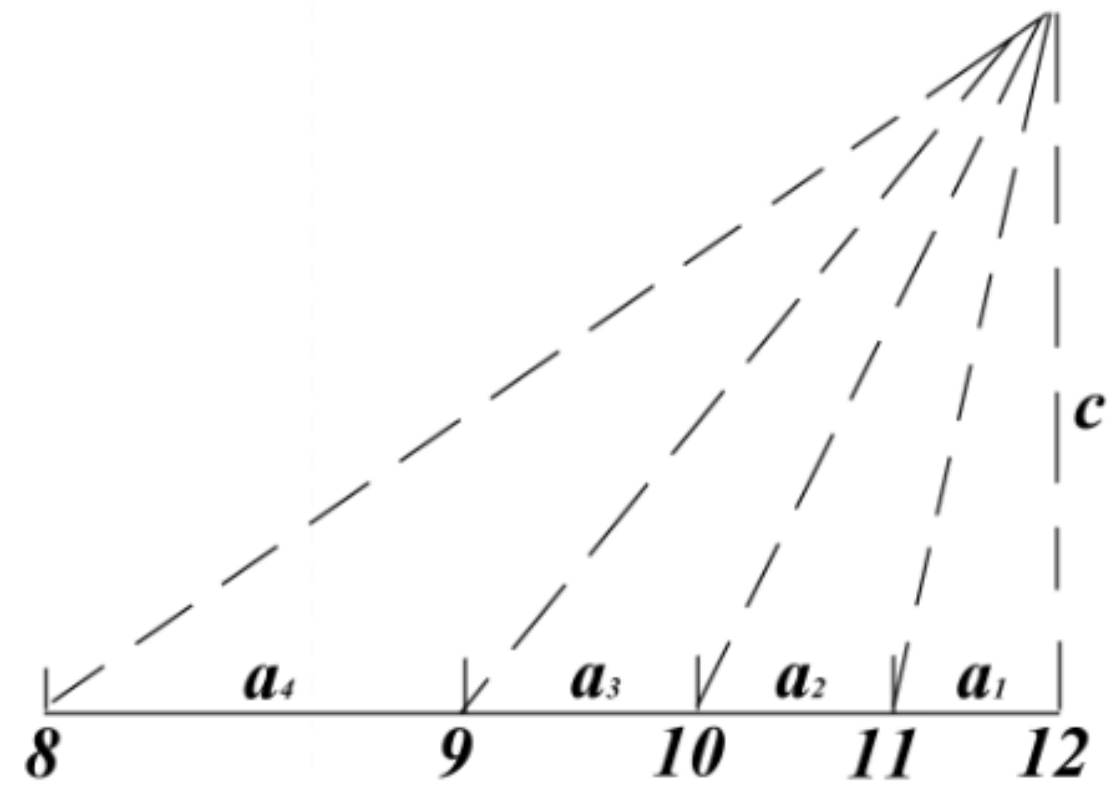

**Figure 9.** Scheme illustrating the location of segments ai formed by the intersection of adjacent hour lines from 8 to 12 o'clock with a line perpendicular to the noon line, *c* - the distance from the place of attachment of the gnomon to the line perpendicular to the noon line.

**Table 3.** Calculated length of segments $a_i$ relative to $a_1$ for a vertical sundial with an inclined gnomon for latitude $\varphi$=25°44'*N*: $Hi^/$ - angle between the noon line and the *i*-th hour line, calculated by formula 1, $t_i$ - time corresponding to *i*-th hour line, ai is the length of the segment bounded by (*i*-1)-th and *i*-th hour lines, and corresponding to the *i*-th hour line.

| *i* | 1 | 2 | 3 | 4 |
|---|---|---|---|---|
| $t_i$, (hour) | 11 | 10 | 9 | 8 |
| $H_i^/$, (°) | 13,6 | 27,5 | 42 | 57,3 |
| $tg(H_i^/)$ | 0,24 | 0,52 | 0,90 | 1,56 |
| $a_i/a_1$ | 1,00 | 1,15 | 1,57 | 2,72 |

Similar calculations were carried out for the hour markings on the tiles found in the Valley of the Kings (Tab. 4).

**Table 4.** Calculated relative length of segments ai for sundials from the Valley of the Kings: $H_i^{/}=H_{f\,i}$ - measured angle between the noon line and the *i*-th hour line, $t_i$ - time corresponding to the *i*-th hour line, $a_i$ - segment length bounded by (*i*-1)-th and *i*-th hour lines, and corresponding to the *i*-th hour line.

| $i$ | 1 | 2 | 3 | 4 |
|---|---|---|---|---|
| $t_i$, (hour) | 11 | 10 | 9 | 8 |
| $H^{/}$,(°) | 14 | 30 | 43 | 57 |
| $tg(H_i^{/})$ | 0,25 | 0,58 | 0,93 | 1,54 |
| $a_i/a_1$ | 1,00 | 1,36 | 1,47 | 2,51 |

Taking into account the linear parameters of the L-shaped sundial from the Berlin Museum, the distances between hour marks were determined from photographs (Table 5), and the relative lengths of segments ai between adjacent hour marks were calculated (Table 6).

**Table 5.** Linear parameters of the L-shaped sundial: $a_i$ – distance between "*i-1*" and "*i*" hour markers.

| Museum number of sundials | Distance from the beginning of the bar to the first mark, (cm) | $a_1$, (cm) | $a_2$, (cm) | $a_3$, (cm) | $a_4$, (cm) | Distance from the last mark to the edge of the bar, (cm) | Total length of the bar [4,5], (cm) |
|---|---|---|---|---|---|---|---|
| №19743 | 3,2 | 2,5 | 3,8 | 5,1 | 6,3 | 9,4 | 30,3 |
| №19744 | 2,9 | 2,4 | 3,4 | 4,3 | 5,7 | 4,5 | 23,2 |

**Table 6.** Lengths of segments of a sundial: $a_i/a_1$ - relative length of the segment corresponding to the *i*-th hour line for a vertical sundial or relative length of the segment between adjacent marks on L-shaped sundial. *I* - vertical sundial with an inclined gnomon for latitude $25^0 44^{/}$ *N* (Valley of the Kings), *II* - vertical sundial found in the Valley of the Kings, *III* - L-shaped sundial № 19743, *IV* - L-shaped sundial № 19744.

| | $a_1/a_1$ | $a_2/a_1$ | $a_3/a_1$ | $a_4/a_1$ |
|---|---|---|---|---|
| *I* | 1,00 | 1,15 | 1,57 | 2,72 |
| *II* | 1,00 | 1,36 | 1,47 | 2,51 |
| *III* | 1,00 | 1,52 | 2,04 | 2,52 |
| *IV* | 1,00 | 1,42 | 1,79 | 2,38 |

For comparison, we also calculated the absolute length of the first segment for a vertical sundial at the level of its lower edge using formula 3:

$$l_1 = c \cdot tg\left(H_1^{/}\right) \tag{3}$$

[4] http://members.aon.at/sundials/berlin-egypte.htm#1.2 (accessed on 01.08.2014)

[5] http://www.aegyptisches-museum-berlin-verein.de/c31.php (accessed on 01.08.2014)

where c=11,8 (cm) - the height of the gnomon attachment relative to the lower edge of the tile from the Valley of the Kings (the length of the noon line), $H_1^{/}$ is the angle of 11 o'clock (between the hour lines of 12 and 11 o'clock).

As a result of calculations according to formula 4, for a latitude of $25°44^{/}$ *N*, we obtain $l_1$=2,86 (cm) for a vertical sundial with an inclined gnomon and $l_1$=2,94 (cm) for marking a sundial from the Valley of the Kings. When taking into account the height of the L-shaped sundial bar, which, in the case of a clock from the Berlin Museum, is approximately equal to one inch - the ancient Egyptian unit of length one "djeba", equal to 1,875 (cm) [22], for vertical sundial we get $l_1$=2,4 (cm) (as for sundial № 19744), and for sundial from the Valley of the Kings $l_1$=2,5 (cm) (as for sundial № 19743).

All these results support the hypothesis of the combined use of the Egyptian L-shaped sundial and vertical sundial with an oblique gnomon. Figure 10 illustrates this combination with a sundial found in the Valley of the Kings and an L-shaped sundial from the Berlin Egyptian Museum.

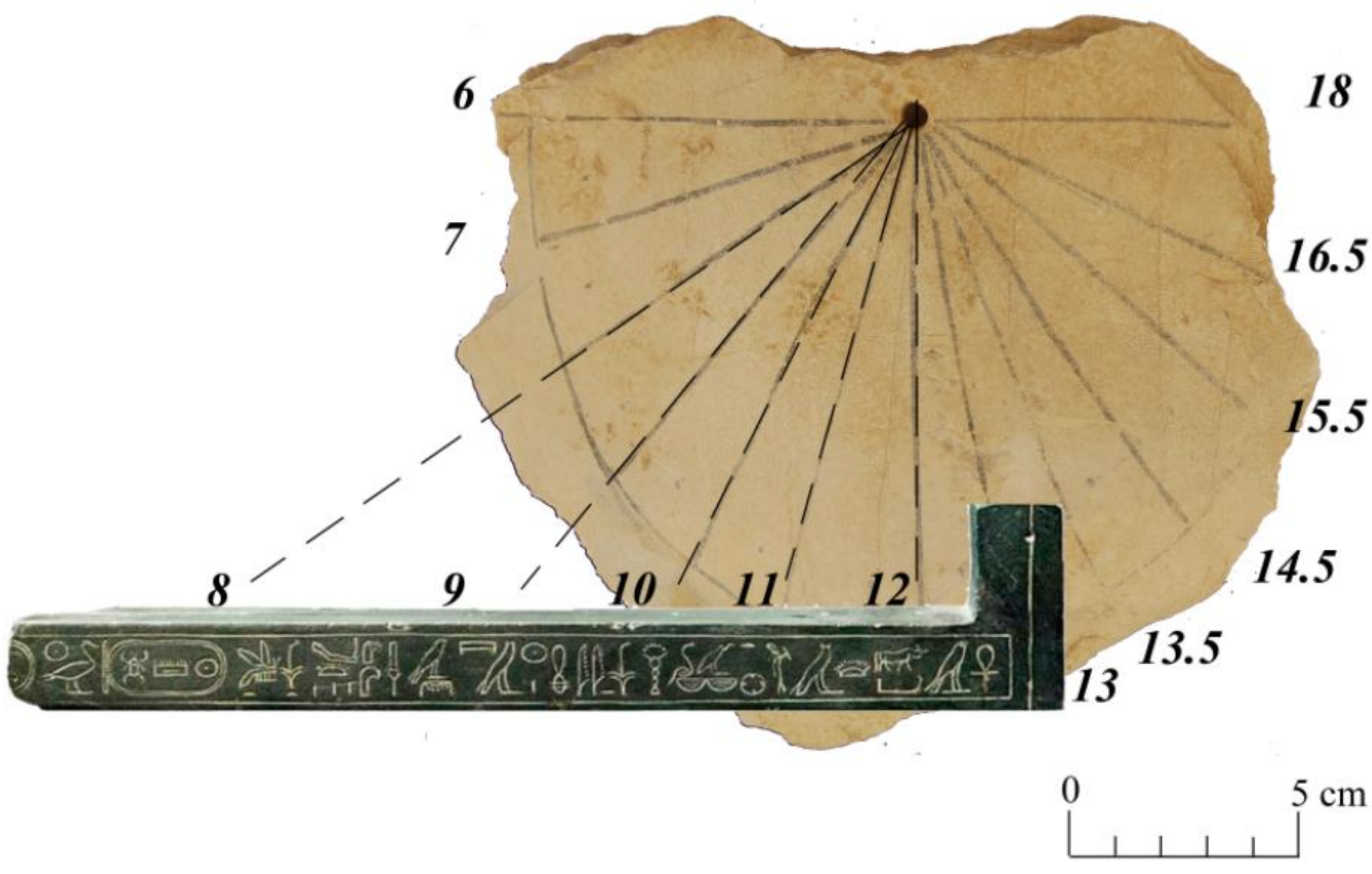

**Figure 10.** Complex of vertical sundial with inclined gnomon, found in the Valley of the Kings, and L-shaped sundial № 19744.

The L-shaped sundial complements the vertical sundial by making it possible to read the inscriptions near the hour markers and interpret the readings of the vertical sundial accordingly, since there are no inscriptions on the vertical sundial. Perhaps such a complex use of two types of sundials was conceived in order to limit the circle of people able to determine the time during the day, including the start and end times of work during the day, by vertical sundials.

In the tomb of Seti I in Abydos, above the image of an L-shaped sundial, next to the marks, there are signatures in the form of Egyptian numbers: 3, 6, 9, 12 (Fig. 6). Until now, there is no satisfactory hypothesis to explain how a sundial with such markings measured time, and what these numbers actually meant.

In the course of this study, it was suggested that these numbers could be related to the relative lengths of the segments between adjacent marks bi in the case when the countdown starts from the noon line, and the first hour line corresponds to 12,5 o'clock, the second to 13,5 o'clock, etc. (Fig. 11). The calculation was carried out according to formula 4, similar to formula 2:

$$b_i/b_1 = \left(tg(H_i^/) - tg(H_{i-1}^/)\right) \Big/ tg(H_1^/) \quad (4)$$

where $H_i^/$ - hour angle between the noon line and the *i*-th hour line, $H_1^/$ - hour angle between the noon line and the first hour line, *i* - the number of the hour line, where *i*=1 for the hour line at 12,5 o'clock, and *i*=4 for 15,5 o'clock; $b_i$ - the length of the segment bounded by (*i*-1)-th and *i*-th hour lines, corresponding to the *i*-th hour line.

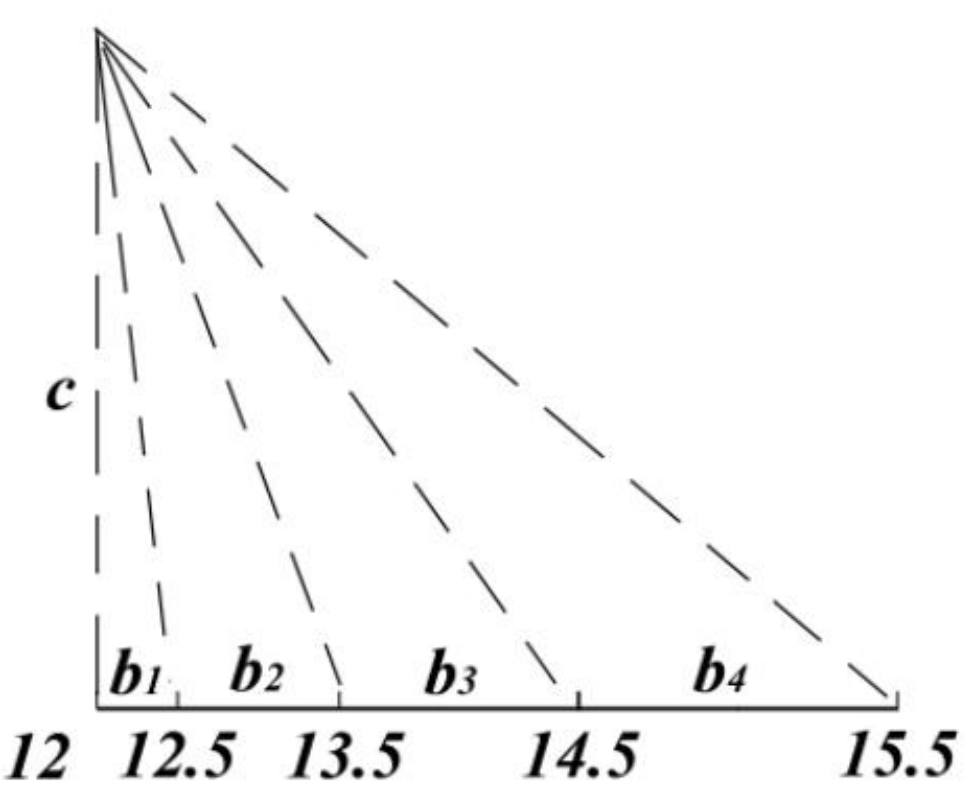

**Figure 11.** Scheme illustrating the location of the segments $b_i$ formed by the intersection of adjacent hour lines from 12 to 15,5 o'clock with a line perpendicular to the noon line, *c* - the distance from the place of attachment of the gnomon to the line perpendicular to the noon line.

The results of calculations by formula 5 for a vertical sundial with an oblique gnomon at latitude $25^0 44^/$ *N* (Valley of the Kings, Egypt), in the time range from 12 to 15,5 hours are presented in Table 7.

The calculated relative lengths of the segments bi, when rounded to integers and increased by a factor of three, give the same series of numbers as in the fresco in the tomb of Seti I: 3, 6, 9, 12. Since these numbers have a common factor, it is logical to assume that the fresco depicted not the relative lengths of the segments, but their absolute values, for example, the lengths in ancient Egyptian inches (djeba). Thus, a series of numbers on a fresco in the tomb of Seti I also confirms the existence of an Egyptian sundial with hour markings shifted by half an hour after 12 noon.

**Table 7.** Calculated length of segments $a_i$ relative to $a_1$ for a vertical sundial with an inclined gnomon for latitude $\varphi$=25°44'*N*: $Hi^{/}$ - angle between the noon line and the *i*-th hour line, calculated by formula 1, $t_i$ - time corresponding to *i*-th hour line, *ai* - the length of the segment bounded by (*i*-1)-th and *i*-th hour lines, and corresponding to the *i*-th hour line.

| *i* | 1 | 2 | 3 | 4 |
|---|---|---|---|---|
| $t_i$, (hour) | 12,5 | 13,5 | 14,5 | 15,5 |
| $H^{/}$,( $^{0}$) | 6,8 | 20,5 | 34,7 | 49,6 |
| $tg(H_i^{/})$ | 0,12 | 0,37 | 0,69 | 1,17 |
| $b_i$ | 1,0 | 2,1 | 2,7 | 4,0 |
| $\approx b_i$ | 1 | 2 | 3 | 4 |
| $\approx b_i \times 3$ | 3 | 6 | 9 | 12 |

The proposed method of measuring time with the help of an L-shaped sundial does not contradict the instructions for their use, placed on a fresco in the tomb of Seti I under the drawing of a sundial, the translation of which was completed and published in part [23].

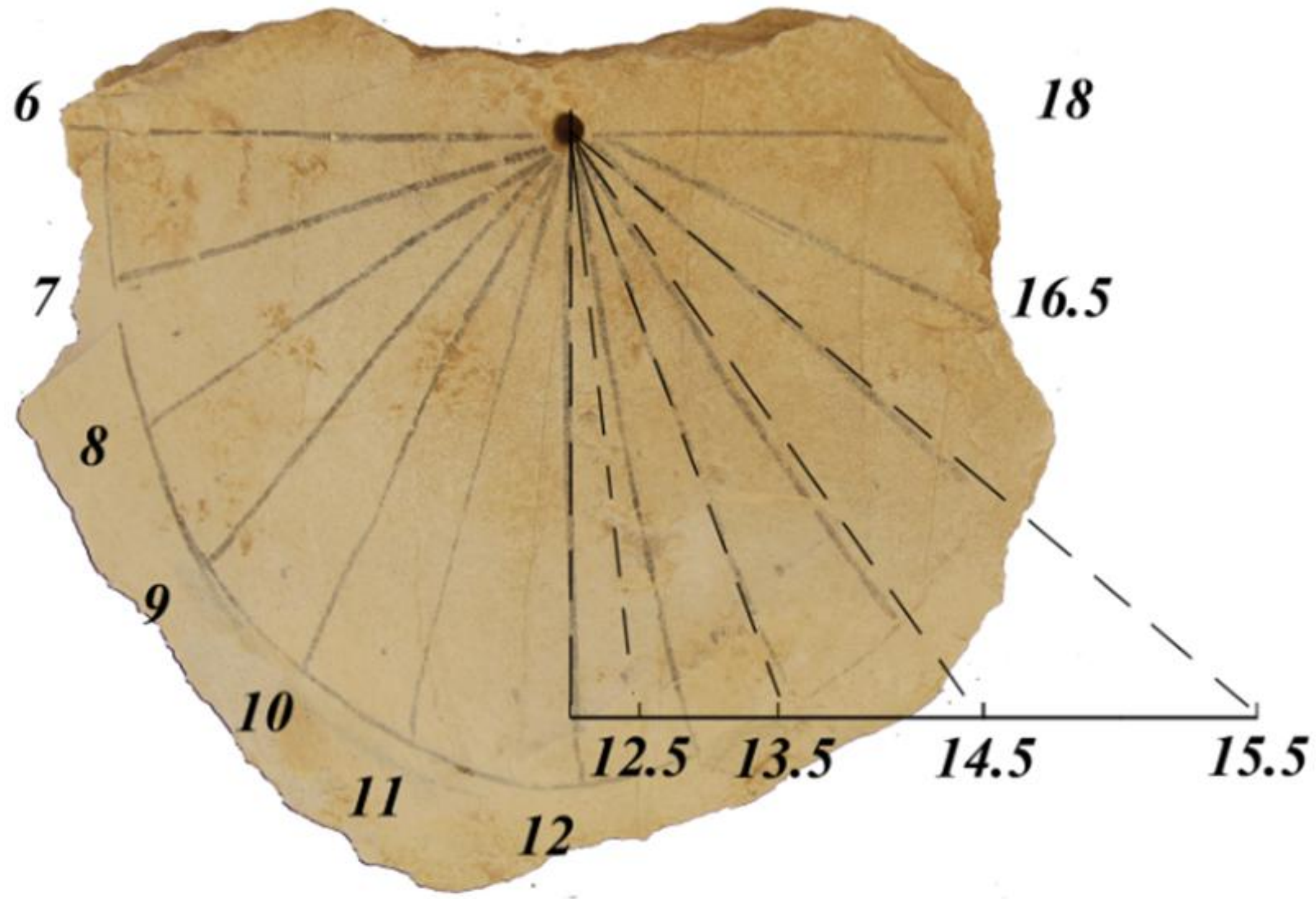

**Figure 12.** Complex of a vertical sundial with an inclined gnomon found in the Valley of the Kings and an L-shaped sundial with hour markings shifted by half hour after 12 noon. The dotted lines illustrate the directions of the gnomon's shadow corresponding to the hour markers on the L-shaped sundial.

Figure 12 illustrates the combined use of a vertical sundial (using the Valley of the Kings sundial as an example) and an L-shaped sundial with afternoon hour markers shifted by half an hour.

It seems that L-shaped sundials without hour offsets were used to measure time and determine the start and end of work in the morning, and sundials with a half-hour offset were used to determine lunch and working hours after 12 noon.

**Conclusion**

Thus, as a result of this study, a new interpretation of the hourly markings of the ancient Egyptian sundial with an inclined gnomon, found in the Valley of the Kings, was proposed. It was concluded that the markings on the tile represent the hourly markings of a sundial with an oblique gnomon for the latitude of the Valley of the Kings, offset by half an hour after 12 noon. The half-hour shift could be associated with an hour and a half rest for workers after 12 noon - a traditional siesta, typical for countries with a hot climate.

A similar marking with the allocation of one and a half hours in the afternoon has already been found on the sundial of the XIII - XII centuries BC [24]. Within the framework of our proposed model, the average error of marking a sundial from the Valley of the Kings is ≈1º or ≈4 minutes. Accuracy offered by S. Bickel and R. Gautschy models: for a vertical sundial with an inclined gnomon 3.9º or 15.6 minutes, and with a horizontal gnomon 6.6º or 42.8 minutes. Thus, the accuracy of the reconstructed model of the ancient Egyptian sundial proposed in this article, found in the Valley of the Kings, is noticeably higher than the accuracy of the models proposed in the article by S. Bickel and R. Gautschy, which testifies in favor of the truth of the model described in this article.

The article also presents a reconstruction of the technology of joint time measurement using an L-shaped sundial and a vertical sundial with an inclined gnomon. In the process of research, it was found that there were two types of L-shaped sundial: with hourly markings without offset before noon and with a shift of half an hour after noon. The L-shaped sundial of these two types supplemented the vertical sundial by allowing the inscriptions next to the hour lines to be read.

References

1. Antoniadi E.M. L'astronomie eguptienne: Depuis les temps les plus recuiles jusqu'a la fin de 1epoque Alexandrine. P., 1934.
2. Kann G. Kratkaya istoriya chasovogo iskusstva [Brief history of watchmaking]. L., 1926.
3. Pipunyrov V.N. Istoriya chasov s drevnejshix vremen do nashix dnej [The history of watches from ancient times to the present day]. Moscow: Nauka, 1982, p. 21.
4. Breamsted J.H. The beginnings of time measurement and the origins of our calendar.— In: Time and its mysteries . Ser. 1.N.Y.; L., 1936.
5. Bikerman E. Xronologiya drevnego mira. Blizhnij Vostok i antichnost'. [Chronology of the ancient world. Near East and antiquity]. - M., 1976 - P. 11.
6. Borchardt L. Agyptische Zeitmessung, 1920
7. Frankfort H. The Cenotaph of Seti I at Abydos (Egypt Exploration Society. Excavation Memoirs 39), London, 1933, vol.2, plate 33.
8. Borchardt L. Altägyptische Sonnenuhren, ZÄS, 1911, vol. 48, pp. 9-17.
9. Pipunyrov V.N. Istoriya chasov s drevnejshix vremen do nashix dnej [The history of watches from ancient times to the present day]. Moscow: Nauka, 1982, fig. 5.
10. Bickel S., Gautschy R. Eine ramessidische Sonnenuhr im Tal der Könige. Zeitschrift für Ägyptische Sprache und Altertumskunde 2014; 141(1): 3–14.
11. Bickel S., Gautschy R. Eine ramessidische Sonnenuhr im Tal der Könige. Zeitschrift für Ägyptische Sprache und Altertumskunde 2014; 141(1): 3–14, Abb. 6.
12. Bickel S., Gautschy R. Eine ramessidische Sonnenuhr im Tal der Könige. Zeitschrift für Ägyptische Sprache und Altertumskunde 2014; 141(1): 3–14.
13. Rohr RRJ. Sundials: History, Theory, and Practice. New York: Dover, 1996, p. 65.
14. Savoie D. Sundials design construction and use. Springer, 2009, fig. 7.2.
15. Bickel S., Gautschy R. Eine ramessidische Sonnenuhr im Tal der Könige. Zeitschrift für Ägyptische Sprache und Altertumskunde 2014; 141(1): 3–14, Abb. 1.
16. Bickel S., Gautschy R. Eine ramessidische Sonnenuhr im Tal der Könige. Zeitschrift für Ägyptische Sprache und Altertumskunde 2014; 141(1): 3–14, Abb. 1.
17. Ideler L. Lehrbuch der Chronologie. Publisher: A. Rücker, 1831, p. 260.
18. Frankfort H. The Cenotaph of Seti I at Abydos (Egypt Exploration Society. Excavation Memoirs 39), London, 1933, vol.2, plate 33.
19. Borchardt L. Altägyptische Sonnenuhren, ZÄS, 1911, vol. 48, pp. 9-17.
20. Symons S. Ancient Egyptian astronomy: timekeeping and cosmography in the new kingdom. University of Leicester, 1999, p. 133.

21. Griffith F. Ll. Flinders P. W. M., Brugsch H. Two Hieroglyphic Papyri from Tanis. London, Trübner & Co., 1889, pl. XV.
22. Clagett M. Ancient Egyptian Science: Ancient Egyptian mathematics. American Philosophical Society, 1999, pp. 7-8.
23. Symons S. Ancient Egyptian astronomy: timekeeping and cosmography in the new kingdom. University of Leicester, 1999, p. 130.
24. Vodolazhskaya L. Analemmatic and horizontal sundials of the Bronze Age (Northern Black Sea Coast). Archaeoastronomy and Ancient Technologies 2013, 1(1), pp. 68-88.